\documentclass[journal,transmag]{IEEEtran}
\usepackage{times,amsmath,epsfig}
\usepackage{subfigure}
\usepackage{booktabs}
\usepackage{hyperref}
\usepackage{array}
\usepackage{graphicx}
\usepackage{epstopdf}
\usepackage{textcomp}
\usepackage{multirow}
\usepackage{perpage}
\usepackage{helvet}
\usepackage{courier}
\usepackage{amsmath,amssymb,epsfig}
\usepackage{bm}
\usepackage{makecell}
\usepackage{epstopdf}

\begin{document}

\title{Efficient Medical Image Segmentation with Intermediate Supervision Mechanism}

\author{Di Yuan$^{1,2}$, Junyang Chen$^{3}$, Zhenghua Xu$^{1,2,\dag}$, Thomas Lukasiewicz$^4$, Zhigang Fu$^5$, Guizhi Xu$^{1,2}$ 
	\thanks{$^1$State Key Laboratory of Reliability and Intelligence of Electrical Equipment, Hebei University of Technology, China.}
	\thanks{$^2$Tianjin Key Laboratory of Bioelectromagnetic Technology and Intelligent Health, Hebei University of Technology, China.}
	\thanks{$^3$Faculty of Science and Technology, University of Macau, China.}
    \thanks{$^4$Department of Computer Science, University of Oxford, United Kingdom.}
	\thanks{$^5$Department of Health Management Center, 983 Hospital of Joint Logistics Support Force, China.}
	\thanks{$^{\dag}$Corresponding author, email: zhenghua.xu@hebut.edu.cn}}


\maketitle
\begin{abstract}
Because the expansion path of U-Net may ignore the characteristics of small targets, intermediate supervision mechanism is proposed.
The original mask is also entered into the network as a label for intermediate output.
However, U-Net is mainly engaged in segmentation, and the extracted features are also targeted at segmentation location information, and the input and output are different.
The label we need is that the input and output are both original masks, which is more similar to the refactoring process, so we propose another intermediate supervision mechanism.
However, the features extracted by the contraction path of this intermediate monitoring mechanism are not necessarily consistent. For example, U-Net's contraction path extracts transverse features, while auto-encoder extracts longitudinal features, which may cause the output of the expansion path to be inconsistent with the label.
Therefore, we put forward the intermediate supervision mechanism of shared-weight decoder module.
Although the intermediate supervision mechanism improves the segmentation accuracy, the training time is too long due to the extra input and multiple loss functions.
For one of these problems, we have introduced tied-weight decoder.
To reduce the redundancy of the model, we combine shared-weight decoder module with tied-weight decoder module.
\end{abstract}



\IEEEdisplaynontitleabstractindextext

\IEEEpeerreviewmaketitle

\section{Introduction}
Although U-Net~\cite{ronneberger2015u} and its variants have already achieved some great successes, their segmentation accuracies for small objects in medical images are still unsatisfaction.
Specifically, in the context of medical images, the objects of interest are often relatively small, e.g., early tumor lesion~\cite{zhang2017multiregion}.
Moreover, in the down-sampling of U-Net, more and more abstract or coarse feature maps will be generated layer by layer~\cite{hamaguchi2018effective}.
Therefore, in the deepest feature maps, the features of these important small objects may become invisible or even be lost~\cite{fu2015relaxing}, which thus results in inaccurate segmentation for small objects.

Consequently, to solve this problem, we propose the intermediate supervision mechanism to enhance the small objects learning capability of U-Net based deep segmentation models.
Intuitively, we believe this enhancement can be achieved by adding some additional intermediate supervision signals for the coarse small object features~\cite{lee2015deeply}.
We denote the deep model that integrates U-Net with intermediate supervision mechanism as \emph{Inter-U-Net}.

Despite achieving better accuracies than U-Net, Inter-U-Net is very time-consuming in practice, mainly due to the need for computing multiple additional intermediate supervision losses and taking the segmentation masks as additional inputs.
Besides, the learning signals in Inter-U-Net become minuscule and insignificant when they are back-propagated to the first layers, resulting into numerous training epochs needed to reach model convergence.
Therefore, to enhance the training efficiency, we further propose to integrate a tied-weight decoder module with Inter-U-Net (denoted as \emph{TW-Inter-U-Net}).
Our experiment results demonstrate that the tied-weight decoder module can greatly reduce the model's training time-cost, while maintaining similar (and sometimes even better) training quality and segmentation accuracies.

Please note that the above intermediate supervision mechanism and tied-weight decoder module can be applied to all U-Net based deep segmentation models. Therefore, as an example, we further illustrate the way of integrating them with a state-of-the-art U-Net variant, U-Net$^{++}$.

The contributions of this paper are briefly as follows:

$\bullet$ We discover the limitation of the existing U-Net based deep models in small objects segmentation, and propose a novel intermediate supervision mechanism to resolve this problem by using additional intermediate supervision signals to help strengthen the coarse small object features and prevent information loss in deep layers.

$\bullet$ A tied-weight decoder module is further introduced to enhance the deep model's training efficiency, while maintaining similar (and sometimes even better) training quality and segmentation accuracies.

$\bullet$ Besides U-Net, we demonstrate that the proposed intermediate supervision mechanism and tied-weight decoder can also be applied in other U-Net based deep segmentation models, e.g., U-Net$^{++}$.

$\bullet$ Extensive experiments are conducted on three real-world datasets, the results show that our proposed method can not only significantly outperform the state-of-the-art baselines in small objects segmentation, but also provide efficient model explanations.



\begin{figure*}[!t]
\centering
\includegraphics[scale=.55]{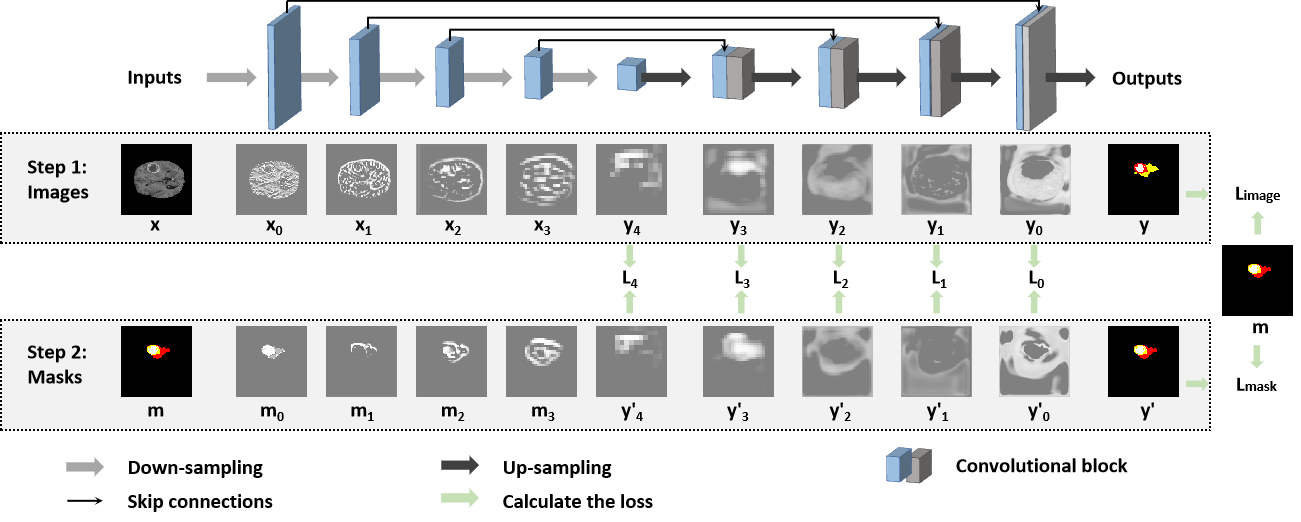}
\caption{Overview of Inter-U-Net.}
\label{Inter-U-Net}
\end{figure*}

\section{Related Work}
\label{sec:RelatedWork}

Deep learning has already been successfully applied in medical image segmentation.
However, most models encounter the problem of poor segmentation performance of small objects.
Therefore, we propose the intermediate supervision mechanism to enhance the small objects learning capability of deep learning based segmentation models.
Besides, to ensure the scalability in practice, we further introduce a tied-weight decoder module to improve our model's training efficiency.

There also exists some other works that utilize additional supervision signals to improve the performance of deep learning models.
Specifically, DSN~\cite{lee2015deeply} is proposed that simultaneously minimizes classification error and improves the directness and transparency of the hidden layer learning process.
However, different from our work, the additional supervision signals in these solutions are not used to solve the problem of inaccurate segmentation of medical images, nor the problem of inaccurate segmentation of small objects.
Besides, in order to enhance the model's learning efficiency, Myronenko et al.~\cite{myronenko20183d} add a variational auto-encoder branch to the encoder endpoint to reconstruct the original image to regularize the tied decoder and impose additional constraints on its layers.
However, these works are based on the variational decoder, and we introduce the tied-weight decoder module.
We believe that the tied-weight decoder module is better than the variational decoder for improving model efficiency.


\section{Inter-U-Net}
\label{sec:Inter-U-Net}

\begin{figure*}[!t]
\centering
\includegraphics[scale=.55]{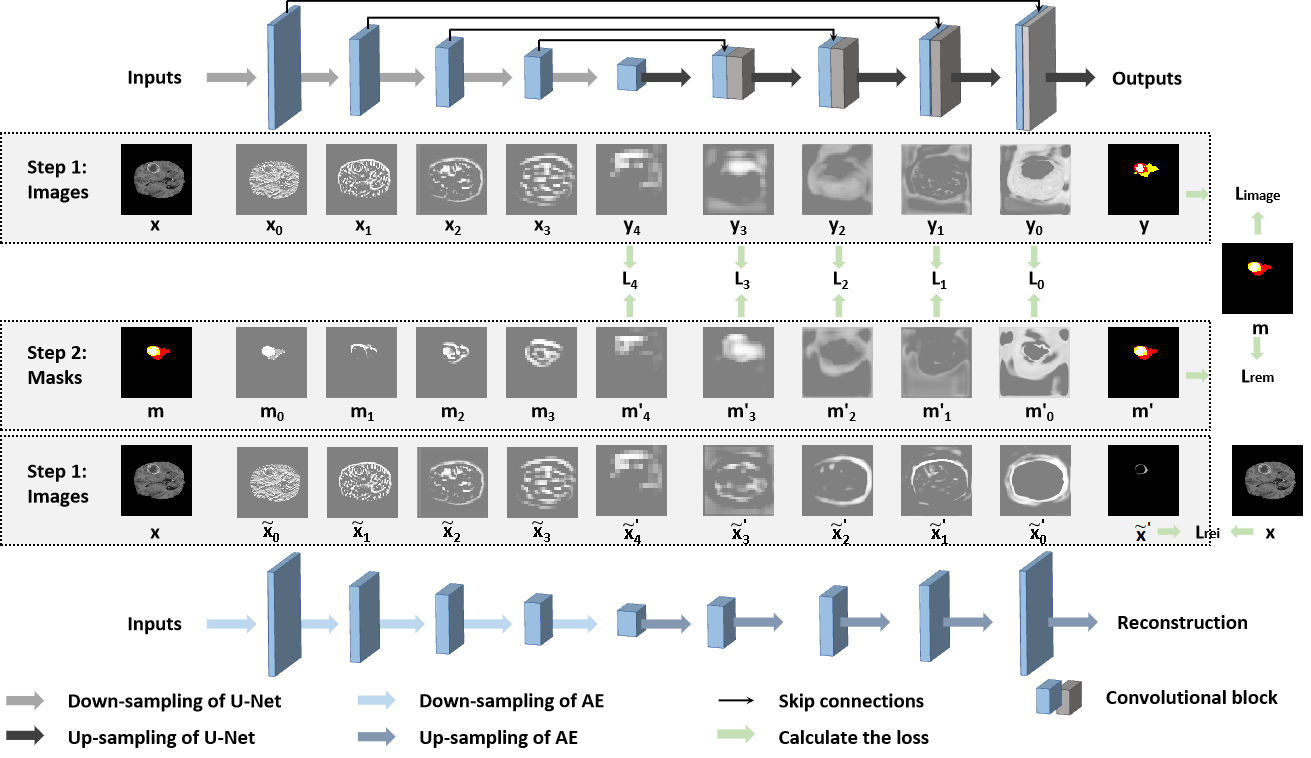}
\caption{Overview of AE-U-Net.}
\label{AE-U-Net}
\end{figure*}

Although U-Net~\cite{ronneberger2015u} has already achieved some good successes in medical image segmentation, the segmentation accuracies for small objects are still unsatisfaction.
In the contracting path of U-Net, the features of these important small objects become less and less visible or even disappear, leading to inaccurate segmentation for the small objects~\cite{fu2015relaxing}.
Consequently, in this paper, we propose five different intermediate supervision mechanisms to overcome the problem of existing U-Net and its variants, and to enhance the segmentation performance of them for medical images with small objects.
Intuitively, we believe this enhancement can be achieved by adding some additional intermediate supervision signals for the abstract or coarse small object feature maps.

In this section, we propose the first intermediate supervision mechanism based on U-Net (\emph{Inter-U-Net}).
Figure~\ref{Inter-U-Net} shows an overview of Inter-U-Net.
Briefly, Inter-U-Net takes the original medical images and the corresponding segmentation masks, $x$ and $m$, as inputs and generate the segmentation outputs at the last layer, denoted $y$ and $y'$.
Besides, $x$ and $m$ generate the intermediate outputs and intermediate masks respectively at the model's deepest layer and the expanding path, denoted $y_j$ and $y'_j$ (where $j=\{0,1,2,3,4\}$), respectively.
For the sake of description, the intermediate outputs and intermediate masks in this paper refer to the outputs produced by the deepest layer of the model and each layer of the actual expansion path.
Finally, the loss function between $y$ and the corresponding segmentation masks $m$ is denoted as $L_{image}$, and the loss function between $y'$ and $m$ is denoted as $L_{mask}$.
Moreover, we regard $y'_j$ as the segmentation masks corresponding to the $j$th layer of $y_j$, and the intermediate supervision losses between  $y_j$ and $y'_j$ are denoted as $L_j$.

Formally, given the original medical images $x$, the corresponding segmentation masks $m$, and their outputs in the last layer of the model, denoted as $y$ and $y'$, the loss functions of them are defined as follows:

\vspace{-1em}
\begin{small}
\begin{align}
L_{image} = L_{BD}(y, m),
\end{align}\label{equ:1}
\end{small}
\vspace{-1.4em}

\vspace{-1.4em}
\begin{small}
\begin{align}
L_{mask} = L_{BD}(y', m),
\end{align}\label{equ:2}
\end{small}
\vspace{-1em}

\noindent where $L_{BD}$ is the combination of Binary Cross-Entropy Loss (BCE Loss) and DICE Loss (also known as F1 score).

Then, the intermediate outputs of $x$ and $m$ of the segmentation model are denoted as $y_j$ and $y'_j$ respectively, and the loss function in the deepest layer of them are denoted as follows:

\vspace{-0.8em}
\begin{small}
\begin{align}
L_{j} = L_M(y_j, y'_j),
\end{align}\label{equ:4}
\end{small}
\vspace{-0.5em}

\noindent where $L_M$ is Mean Square Error (MSE Loss), $j \in \{0,1,2,3,4\}$.
Finally, the hybrid loss function ($L_h^1$) designed by us can be considered as the sum of $L_{image}$, $L_{mask}$, and $L_j$.
Therefore, the model's learning ability for the small objects and the segmentation details in medical images can be enhanced. $L_h^1$ is defined as follows:

\vspace{-0.8em}
\begin{small}
\begin{align}
L_h^1 =\alpha \cdot L_{image} + \gamma \cdot L_{mask} + \lambda \cdot \sum_{j=0} ^4 \omega_j \cdot L_j,
\end{align}
\end{small}
\vspace{-0.5em}

\noindent where $\alpha$, $\gamma$, $\lambda$, and $\omega_i$ respectively represent the weight of each loss function, which are independent parameters that can be adjusted as required.

\section{AE-U-Net}
\label{sec:AE-U-Net}
However, U-Net is mainly engaged in segmentation, and the extracted feature is also the position information for precise positioning, and the input and output of U-NET are different images.
Our intermediate monitoring mechanism requires both the original mask input and output, which is more similar to the reconstruction process of auto-encoder (AE)~\cite{deng2012three}.
Therefore, we combine AE with U-Net and propose the second intermediate supervision mechanism (AE-U-NET). The structure is shown as Figure~\ref{AE-U-Net}.
Different from the first intermediate supervision mechanism, Inter-U-NET directly inputs corresponding segmentation masks into the model, and the intermediate output generated by them is taken as corresponding ground truths, while our AE-U-NET inputs segmentation masks into another auto-Encoder model, and takes the intermediate output generated in AE as corresponding ground truths.
AE-U-Net and its loss function are described as follows.

\section{SAE-U-Net}
\label{sec:SAE-U-Net}

\begin{figure*}[!t]
\centering
\includegraphics[scale=.55]{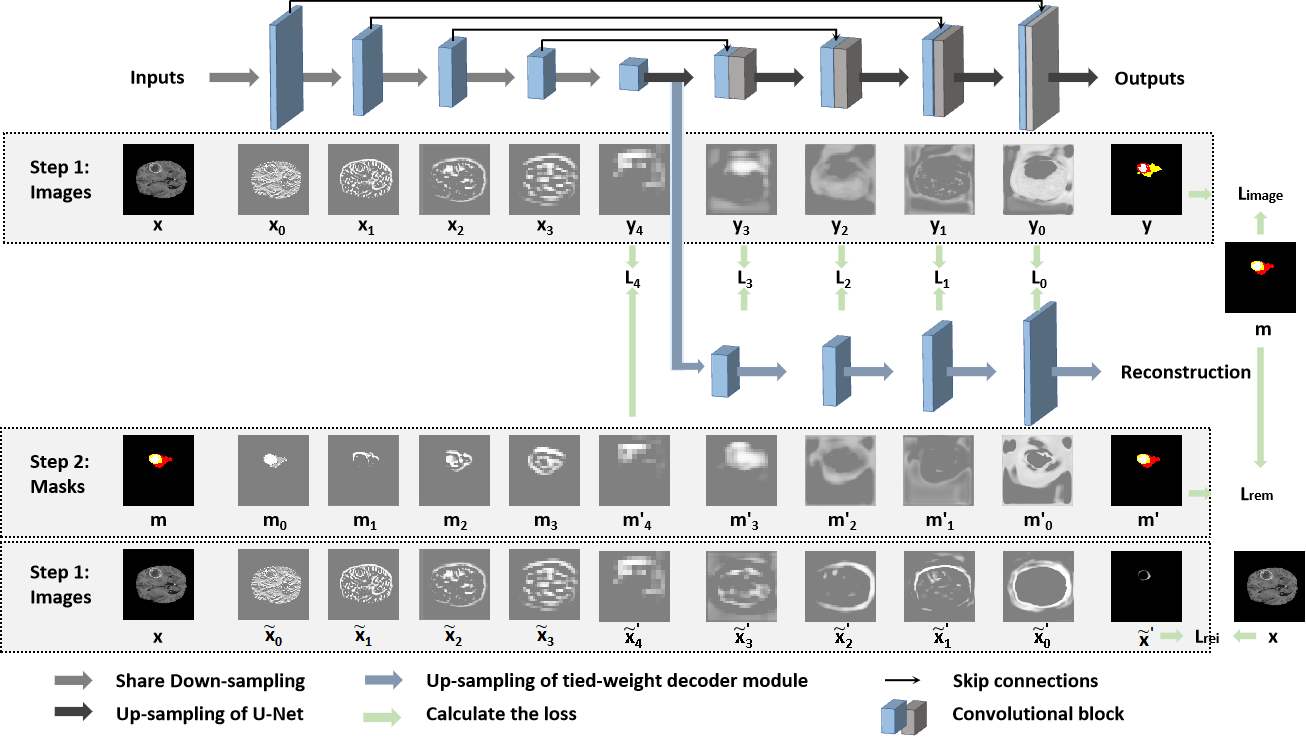}
\caption{Overview of SAE-U-Net.}
\label{SAE-U-Net}
\end{figure*}

However, the features extracted by the contracting path of AE-U-Net are not necessarily consistent. For example, U-Net extracts transverse features in contracting path, while AE extracts longitudinal features, which may cause the output of the expanding path to be inconsistent with the label.
Therefore, we put forward the intermediate supervision mechanism with a shared-weight decoder (\emph{SAE-U-Net}).
The structure is shown as Figure~\ref{SAE-U-Net}.

\begin{figure*}[!t]
\centering
\includegraphics[scale=.55]{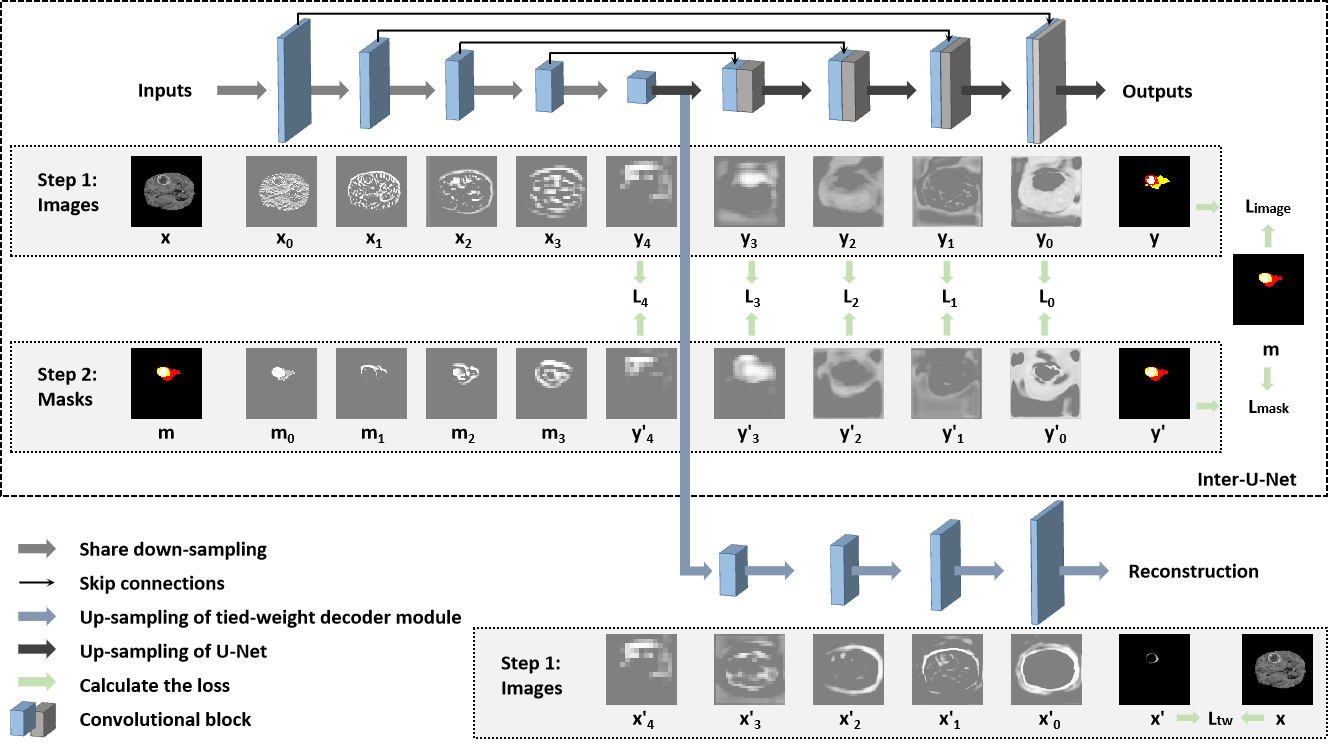}
\caption{Overview of TWI-U-Net.}
\label{TWI-U-Net}
\end{figure*}

\begin{figure*}[!t]
\centering
\includegraphics[scale=.55]{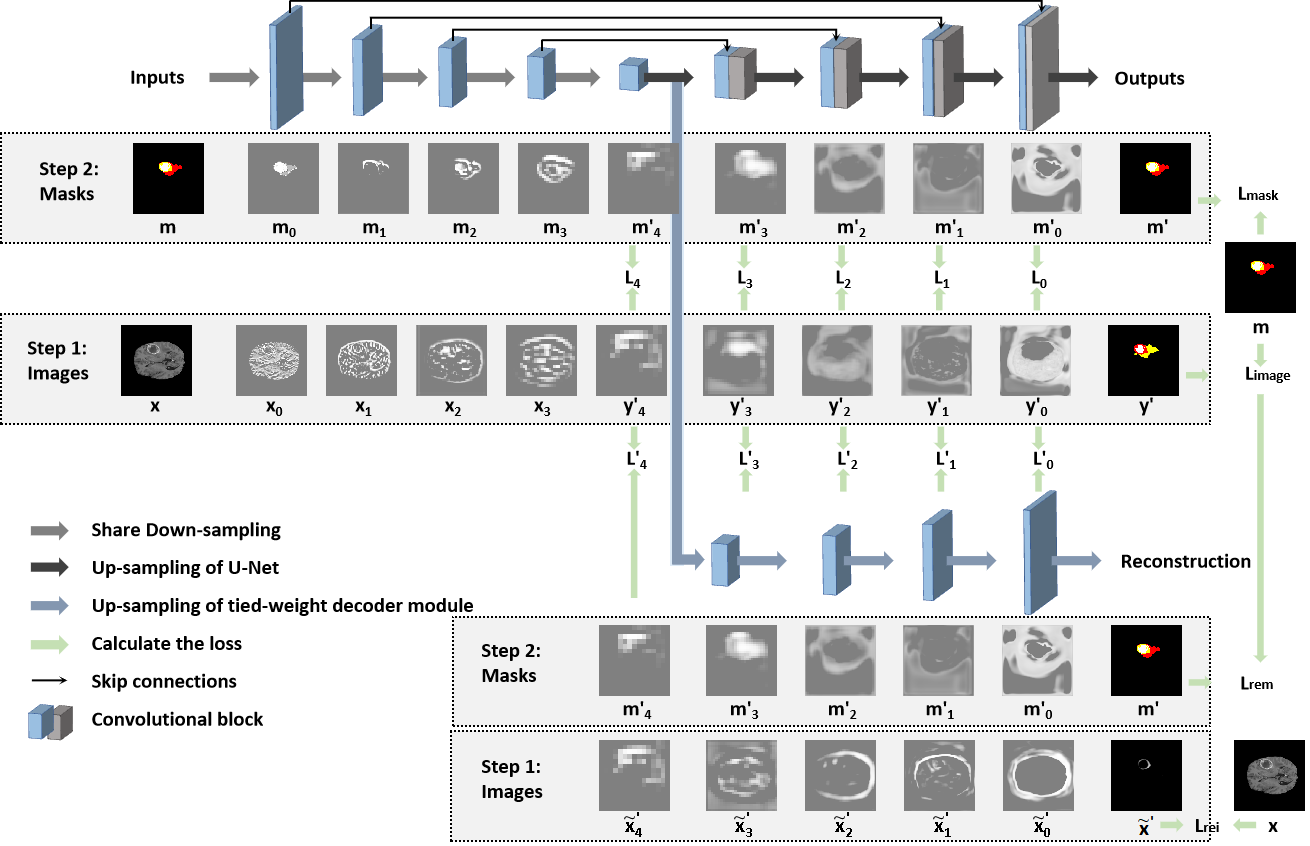}
\caption{Overview of TWAE-U-Net.}
\label{TWAE-U-Net}
\end{figure*}
\section{Tied-weight decoder module}
\label{sec:TW}
Intermediate supervision mechanism can improve the model's segmentation accuracy for the small objects.
Nevertheless, its training is usually very time-consuming, mainly because i) it has additional loss functions and additional inputs.
ii) The learning signals become minuscule and insignificant when they are backpropagated to the first few layers, which results in very slow learning progress and excessive training epochs for model convergence~\cite{xu2017tag}.
Therefore, in this paper, to enhance model's training efficiency, we propose to use a tied-weight decoder module to generate additional learning signals based on reconstruction errors.
The model of Intermediate supervision mechanism with tied-weight decoder module is called \emph{TWI-U-Net}. The structure is shown as Figure~\ref{TWI-U-Net}.
To reduce the redundancy of the model, we combine SAE-U-Net with tied-weight decoder (\emph{TWAE-U-Net}). The structure is shown as Figure~\ref{TWAE-U-Net}.

Formally, the definition of layers in encoder are the same as the ones in Inter-U-Net. As for the decoder, the outputs of the tied-weight decoder module are denoted as $\widetilde{x_j}$ (where $j=\{0,1,2,3,4\}$), and the reconstructed loss function of this module is defined as:

\vspace{-1em}
\begin{small}
\begin{align}
L_t =&\ L_B(\widetilde{x}, x),
\end{align}\label{equ:tw}
\end{small}
\vspace{-0.6em}

\noindent where $L_B$ is BCE Loss. Finally, the reconstruction loss function is combined with the loss function in Inter-U-Net to form an efficiency deep learning signal for model training, and the total loss function is formally defined as:

\vspace{-1em}
\begin{small}
\begin{align}\label{equ:total}
\hspace{-1em}
L_h = L_h^1 + \beta \cdot L_{t},
\end{align}
\end{small}
\vspace{-1em}

\noindent where $\beta$ is an independent variable, which can be adjusted independently as needed.
The experiments show that TW-Inter-U-Net not only achieves much better segmentation performance for the small objects than the state-of-the-art U-Net models, but also has much quicker converging speed in model training.

\section{Experiments}
\label{sec:results}
In this section, we first introduce our experimental implementations, including data preprocessing, baseline models, evaluation metrics, and so on.
Then, we validate the performance of our models with the task of medical image segmentation from both quantitative and qualitative aspects.
Finally, we further evaluate the effectiveness of our models in the model convergence speeds.
Moreover, we also prove that the combination of the intermediate supervision mechanism and the tied-weight decoder module will not weaken the effects of each other, but improve each other.
After that, we also verify that our proposed methods are effective in both U-Net and U-Net variants.

    \subsection{Datasets}
    The empirical studies over three real datasets confirm that our models beat other baseline models.
    We use three medical imaging datasets for model evaluation, covering lesions/organs from different medical imaging modalities. These datasets contain the characteristics of small datasets, small objects, and unbalanced classes, and are more representative of the characteristics of current medical images.
    For all datasets, there are 70\% of the datasets for training, 10\% for validating, and 20\% for testing.

    \subsection{Baseline Models}
    The descriptions of the baseline models can be divided into two groups:
    \emph{U-Net as the backbone,} and
    \emph{U-Net$^{++}$ as the backbone.}

    \subsection{Experiment settings}
    We implement our network in PyTorch and trained it on NVIDIA TITAN XP 12GB GPU using three datasets.
    Our experiments use the \emph{early-stop} mechanism,
    which means the training stops when the model converges or reaches the maximum training epochs.
    Moreover, we use Adam optimizer with an initial learning rate of $\alpha _ 0 = \emph{3e-4}$ and progressively decrease it for every three training epochs according to: $\alpha = \alpha _ 0 \ast 0.9$.

\begin{figure*}[!t]
\centering
\includegraphics[width=15cm]{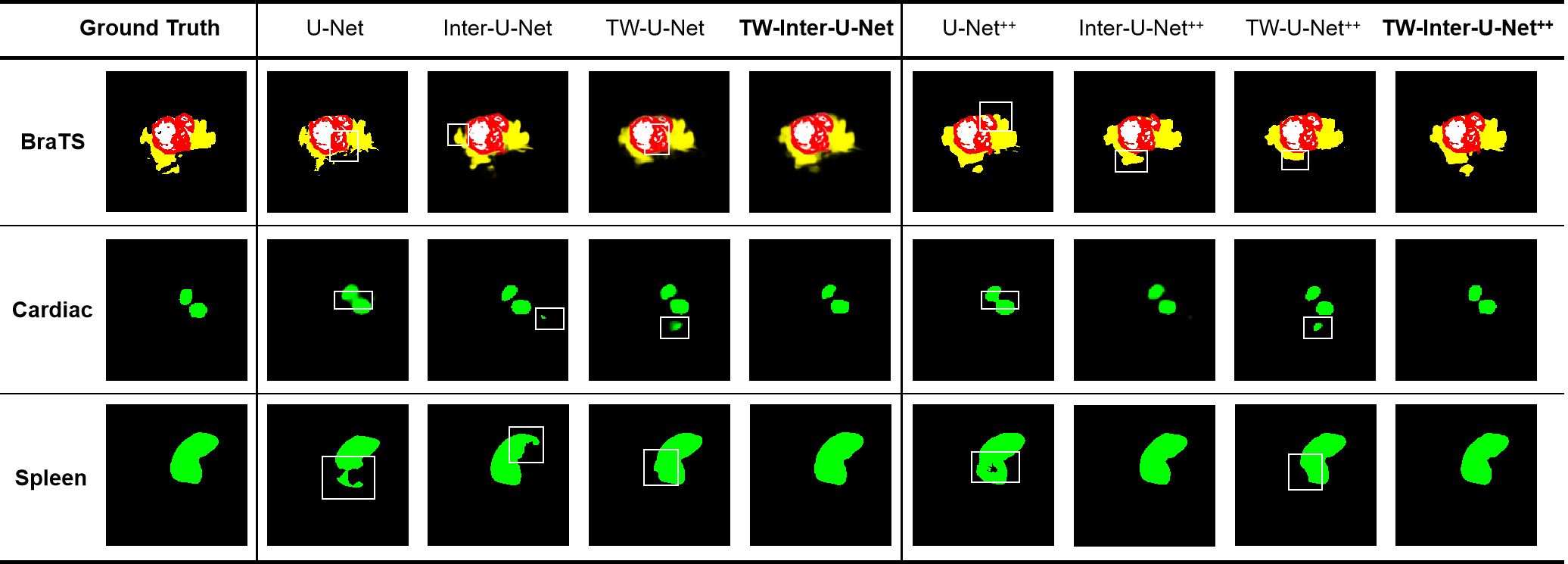}
\caption{\small {Visualization of segmentation results of all models on three datasets.}}
\vspace{0.8em}
\label{resultspng}
\end{figure*}

    \subsection{Main Results}
    We also prove the validity of the proposed method from qualitative.
    Specifically, the first, second, and third rows of Figure~\ref{resultspng} present the results of BraTS (2019), Cardiac, and Spleen datasets, respectively.
    Also, the left and right panels of Figure~\ref{resultspng} denote, respectively, ground truths, U-Net and U-Net-based enhancement models, and U-Net$^{++}$ and its enhancement models.

\section{Conclusions and Future Work}
\label{sec:conclusion}
The segmentation accuracy of U-Net and its variants is not good for small targets.
Therefore, we propose different intermediate supervision mechanisms to solve this problem.
In addition, we introduced tied-weight decoder module to improve the model's training efficiency.

\section*{Acknowledgment}
This work was supported in part by the National Natural Science Foundation of China under grant 61906063,
in part by the Natural Science Foundation of Tianjin, China, under grant 19JCQNJC00400,
in part by the Yuanguang Scholar Fund of Hebei University of Technology, China,
and in part by the National Natural Science Foundation of China under grant 51737003.

\bibliographystyle{elsarticle-num}
\bibliography{ref}

\begin{thebibliography}{1}
\expandafter\ifx\csname url\endcsname\relax
  \def\url#1{\texttt{#1}}\fi
\expandafter\ifx\csname urlprefix\endcsname\relax\def\urlprefix{URL }\fi
\expandafter\ifx\csname href\endcsname\relax
  \def\href#1#2{#2} \def\path#1{#1}\fi

\bibitem{ronneberger2015u}
O.~Ronneberger, P.~Fischer, T.~Brox, U-net: Convolutional networks for
  biomedical image segmentation, in: Proceedings of the International
  Conference on Medical Image Computing and Computer-Assisted Intervention,
  2015, pp. 234--241.

\bibitem{zhang2017multiregion}
J.~Zhang, D.~Su, J.~Fujimoto, L.~Ying, C.-W. Chow, W.~Sun, J.~Zhang, J.~Hu,
  C.~Behrens, M.~Antonoff, et~al., Multiregion whole exome seuquencing of
  pre-and early neoplastic lung lesions, 2017.

\bibitem{hamaguchi2018effective}
R.~Hamaguchi, A.~Fujita, K.~Nemoto, T.~Imaizumi, S.~Hikosaka, Effective use of
  dilated convolutions for segmenting small object instances in remote sensing
  imagery, in: Proceedings of the IEEE Winter Conference on Applications of
  Computer Vision, 2018, pp. 1442--1450.

\bibitem{fu2015relaxing}
J.~Fu, Y.~Wu, T.~Mei, J.~Wang, H.~Lu, Y.~Rui, Relaxing from vocabulary: Robust
  weakly-supervised deep learning for vocabulary-free image tagging, in:
  Proceedings of the IEEE International Conference on Computer Vision, 2015,
  pp. 1985--1993.

\bibitem{lee2015deeply}
C.-Y. Lee, S.~Xie, P.~Gallagher, Z.~Zhang, Z.~Tu, Deeply-supervised nets, in:
  Proceedings of the International Conference on Artificial Intelligence and
  Statistics, 2015, pp. 562--570.

\bibitem{myronenko20183d}
A.~Myronenko, 3d mri brain tumor segmentation using autoencoder regularization,
  in: Proceedings of the International Conference on MICCAI Brainlesion
  Workshop, 2018, pp. 311--320.

\bibitem{deng2012three}
L.~Deng, Three classes of deep learning architectures and their applications: a
  tutorial survey, APSIPA Transactions on Signal and Information Processing,
  2012.

\bibitem{xu2017tag}
Z.~Xu, T.~Lukasiewicz, C.~Chen, Y.~Miao, X.~Meng, Tag-aware personalized
  recommendation using a hybrid deep model, Proceedings of AAAI
  Press/International Joint Conferences on Artificial Intelligence, 2017.

\end{thebibliography}




\end{document}